\documentclass{ws-ijmpcs}

\usepackage{lineno}  

\newcommand{\Ks}{\mathrm{K_S}}
\newcommand{\Kl}{\mathrm{K_L}}

\newcommand{\ppm}{\pi^{+}\!\pi^{-}\!}
\begin{document}

\markboth{W. Krzemien}
{Recent results from KLOE-2}

%
\catchline{}{}{}{}{}
%

\title{Recent results from KLOE-2}

\author{Wojciech Krzemien\footnote{On behalf of the KLOE-2 collaboration.}}
\address{High Energy Physics Division, National Centre for Nuclear Research, \\
05-400 Otwock-\'Swierk, Poland\\
wojciech.krzemien@ncbj.gov.pl}

\maketitle

\begin{history}
\published{Day Month Year}
\end{history}

\begin{abstract}
The most recent results from the KLOE experiment are presented, covering: the measurement of the running fine-structure constant $\alpha_{em}$, 
the Dalitz plot measurement of $\eta \rightarrow \pi^{+}\pi^{-}\pi^{0}$, 
the search of a U boson, tests of discrete symmetries and quantum coherence.
The KLOE-2 Collaboration will take data until mid 2018 aiming to collect 5 fb$^{−1}$ increasing the data set, in order to produce new precision measurements and continue studies of fundamental symmetries and New Physics.

\end{abstract}

\section{Introduction}


The KLOE detector at the DA$\Phi$NE $\phi$-factory, located at the Laboratori Nazionali di Frascati(INFN-LNF) in Frascati, collected data from 2001 to 2006~\cite{DAFNE-KLOE}.
In November of 2014 the new upgraded KLOE-2~\cite{DAFNE-KLOE2,KLOE2_proposal} detector started to acquire data, with the aim of collecting more than $5 \,\mathrm{fb^{-1}}$ by the end of March 2018. 
Until September of 2017 about $4 \,\mathrm{fb^{-1}}$ have been collected. During this period DA$\Phi$NE peak luminosity has been of $2.2 \times 10^{32} \,\mathrm{cm^{-2}s^{-1}}$, and the integrated daily luminosity has been of about $10 \,\mathrm{pb^{-1}}$.

\section{Running of $\alpha_{em}$}

It is well known that the fine-structure constant, $\alpha_{em}$, is a running parameter due to vacuum polarization effects, $\alpha_{em}(q^2) = \frac {\alpha_{em}(0)}{1 - \Delta\alpha(q^2)}$. The correction $\Delta\alpha$ represents the sum of the lepton, the five lightest quarks, and the top quark contributions: $\Delta\alpha(q^2) = \Delta\alpha_{lep}(q^2) + \Delta\alpha^{(5)}_{had}(q^2) +\Delta\alpha_{top}(q^2))$. At low energies the top quark contribution is negligible, the leptonic part is calculated with very high precision in QED, while the hadronic part cannot be calculated with perturbative methods. Instead, it can be evaluated from experimental data by means of the dispersion relation : $\Delta\alpha^{(5)}_{had}(s) = - \frac{\alpha(0) s}{3\pi} \int_{s0}^{\infty} \frac {R_{had}(s')}{s'(s'-s-i\epsilon)}ds'$ , where $R_{had}(s) = \frac{σ(e^+e^- \rightarrow \mathrm{hadrons})}{(e^+e^- \rightarrow \mu^+\mu^-)}$ and $(s = q^2)$. The value of $\alpha_{em}(s)$ in the time-like region can be extracted from the ratio of the differential cross-section of $e^+e^- \rightarrow \mu^+\mu^-\gamma$, with a photon from the Initial State Radiation (ISR), and the corresponding cross-section obtained from the Monte Carlo (MC) simulation with $\alpha_{em}(s) = \alpha_{em}(0)$~\cite{aem-1}.
\begin{equation}
  \centering
  \left|\frac{\alpha_{em}(s)}{\alpha_{em}(0)}\right|^2 = \frac{d\sigma_{data}(e^+e^- \rightarrow \mu^+ \mu^- \gamma (\gamma))|_{ISR}/d\sqrt{s}}{d\sigma^0_{MC}(e^+e^- \rightarrow \mu^+ \mu^- \gamma (\gamma))|_{ISR}/d\sqrt{s}}  
  \label{eq:1}
\end{equation}

\begin{figure}[!]
  \centerline{\includegraphics[width=6cm]{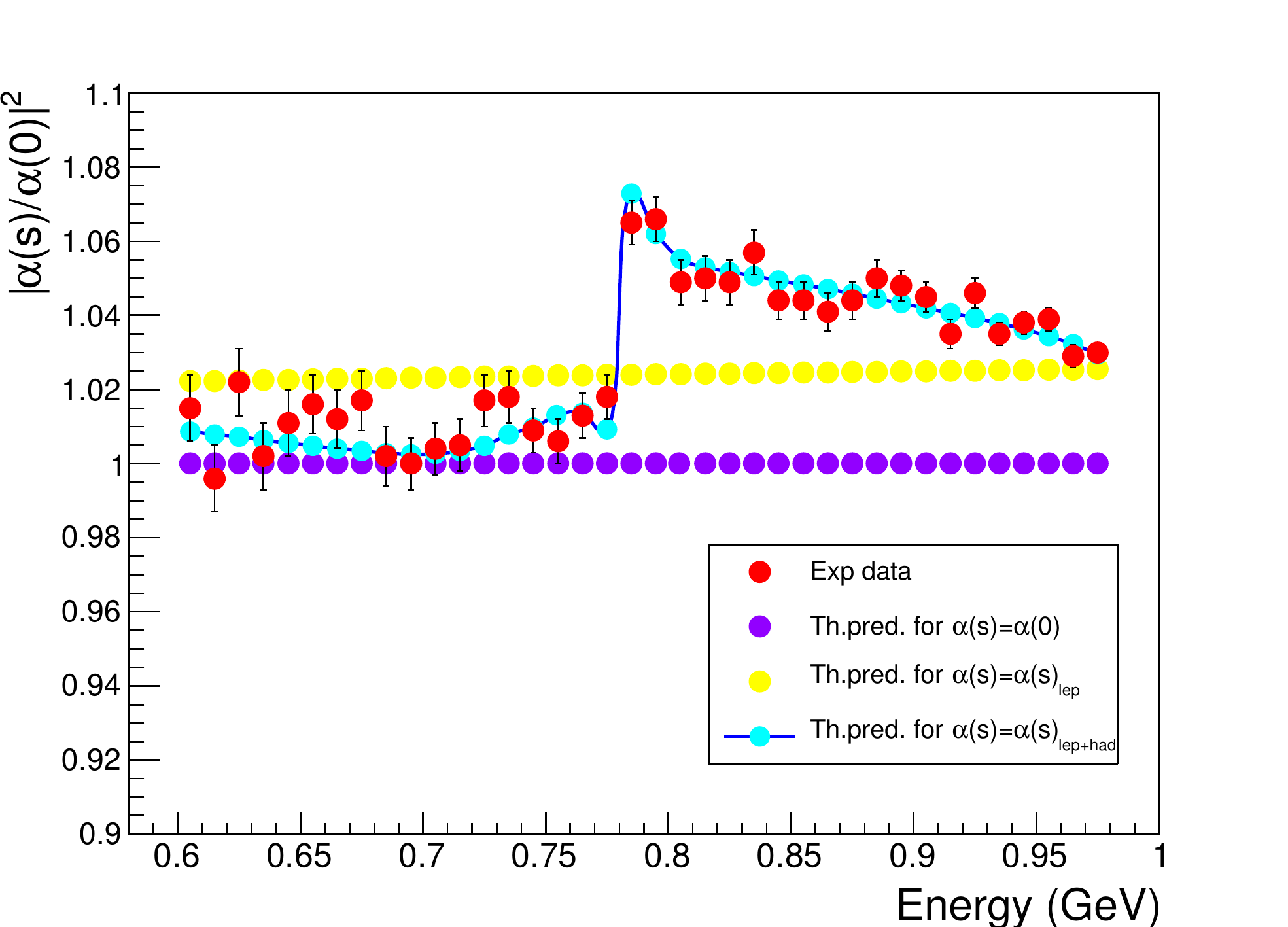}}
  \caption{ $\left|\frac{\alpha_{em}(s)}{\alpha_{em}(0)}\right|^2$ as a function of the di-muon invariant mass compared to the theoretical predictions.}
  \label{fig:aem1}
\end{figure}

In Fig.~\ref{fig:aem1} the ratio  ~\ref{eq:1} is compared to the theoretical predictions~\cite{aem_theory}.
In the time-like region of $q^2$, $\Delta\alpha$ is a complex quantity, the real part of it expressed as:

\begin{equation}
  \centering
  \Re\Delta\alpha= 1 - \sqrt{|\alpha_{em}(0)/\alpha_{em}(s)|^2 - (\Im\Delta\alpha)^2}
  \label{eq:aem2}
\end{equation}

\begin{figure}
  \centering
  \includegraphics[width=.4\linewidth]{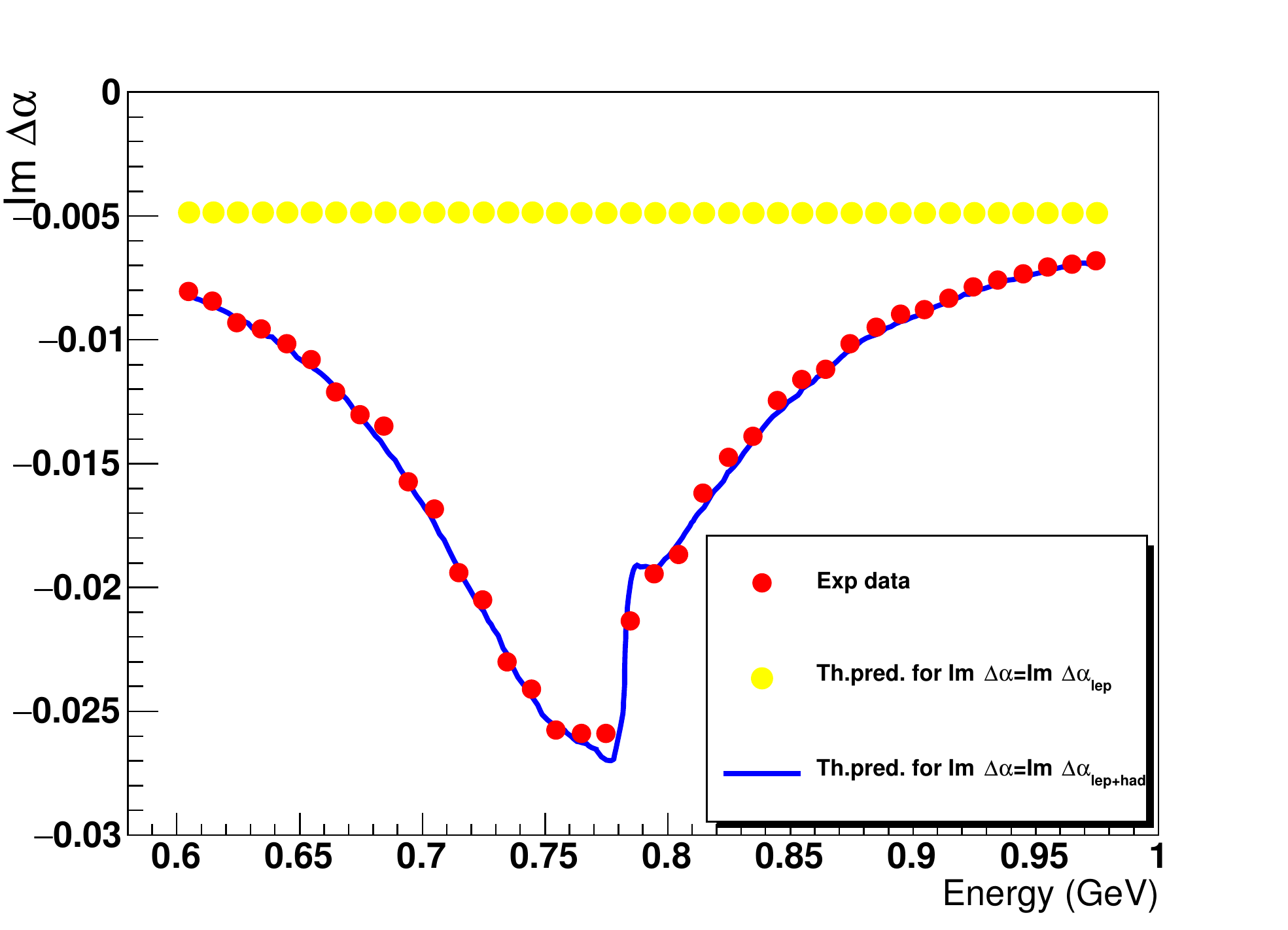}
  \includegraphics[width=.45\linewidth]{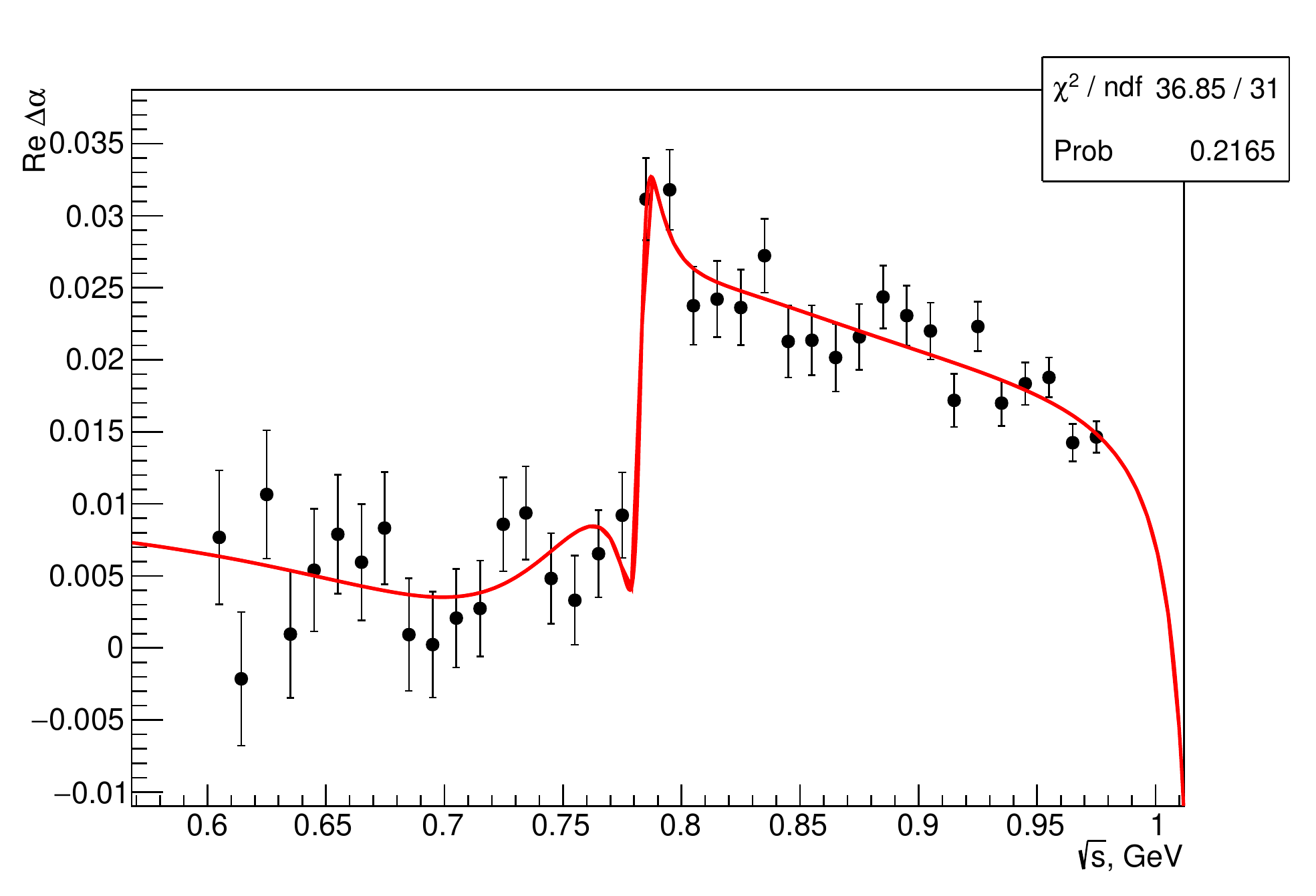}
  \caption{Left: Im$\Delta\alpha$ from KLOE $\pi^+\pi^-\gamma$ cross section (red points) and from a compilation of other measurements (blue curve); right: Re$\Delta\alpha$ from Eq.~\ref{eq:aem2}, the red line is the fit described in the text.}
  \label{fig:aem2}
\end{figure}

From the optical theorem follows that $\Im\Delta\alpha = -\frac{\alpha_{em}}{3}R_{had}(s)$, and Fig.~\ref{fig:aem2} panel left presents the behavior of $\Im\Delta\alpha$ obtained from the KLOE data on the $\pi^+\pi^-$ cross-section~\cite{2picross} that is the dominant contribution in this energy range. The real part from Eq.~\ref{eq:aem2} is shown in Fig~\ref{fig:aem2}, panel right.

The superimposed curve is a fit performed by parametrizing the $\omega$(782) and $\Phi$(1020) as simple Breit-Wigner and describing the $\rho$(770) with the Gounaris-Sakurai parametrization, and adding a non-resonant term. The mass and the width of the $\Phi$, the $\omega$ width as well as the product Br($\Phi \rightarrow e^+e^−$)Br($ \Phi \rightarrow \mu^+\mu^−$) have been fixed to the PDG values~\cite{PDG}. Assuming lepton universality and correcting for phase space, Br($\omega \rightarrow \mu^+\mu^−$) =$(6.6 \pm 1.4 \pm 1.7)\times10^{−5}$ is obtained (PDG: Br($\omega \rightarrow \mu^+\mu^−$) = $(9.0 \pm 3.1)\times10^{−5})$.
The results show more than 5$\sigma$ significance of the hadronic contribution.

\section{Precision measurement $\eta \rightarrow 3\pi$ Dalitz plot distribution}

The isospin-violating decay $\eta \rightarrow 3\pi$ is induce dominantly by the strong interaction via the $u-d$ quark mass difference. Therefore, the decay is a perfect laboratory for testing chiral perturbation theory, ChPT~\cite{ChPT}. The decay amplitude is proportional to $Q^{−2}$, the quadratic quark mass ratio defined as $Q^2 = \frac{m_s^2 - \hat{m}^2}{m_d^2 - \hat{m}^2}$ where $\hat{m} = \frac{1}{2}\left(m_d+m_u \right)$. Thus, the precise experimental distributions could be used directly for the dispersive analyses to determine the $Q$ ratio.

\begin{figure}
  \centering
  \includegraphics[width=.4\linewidth]{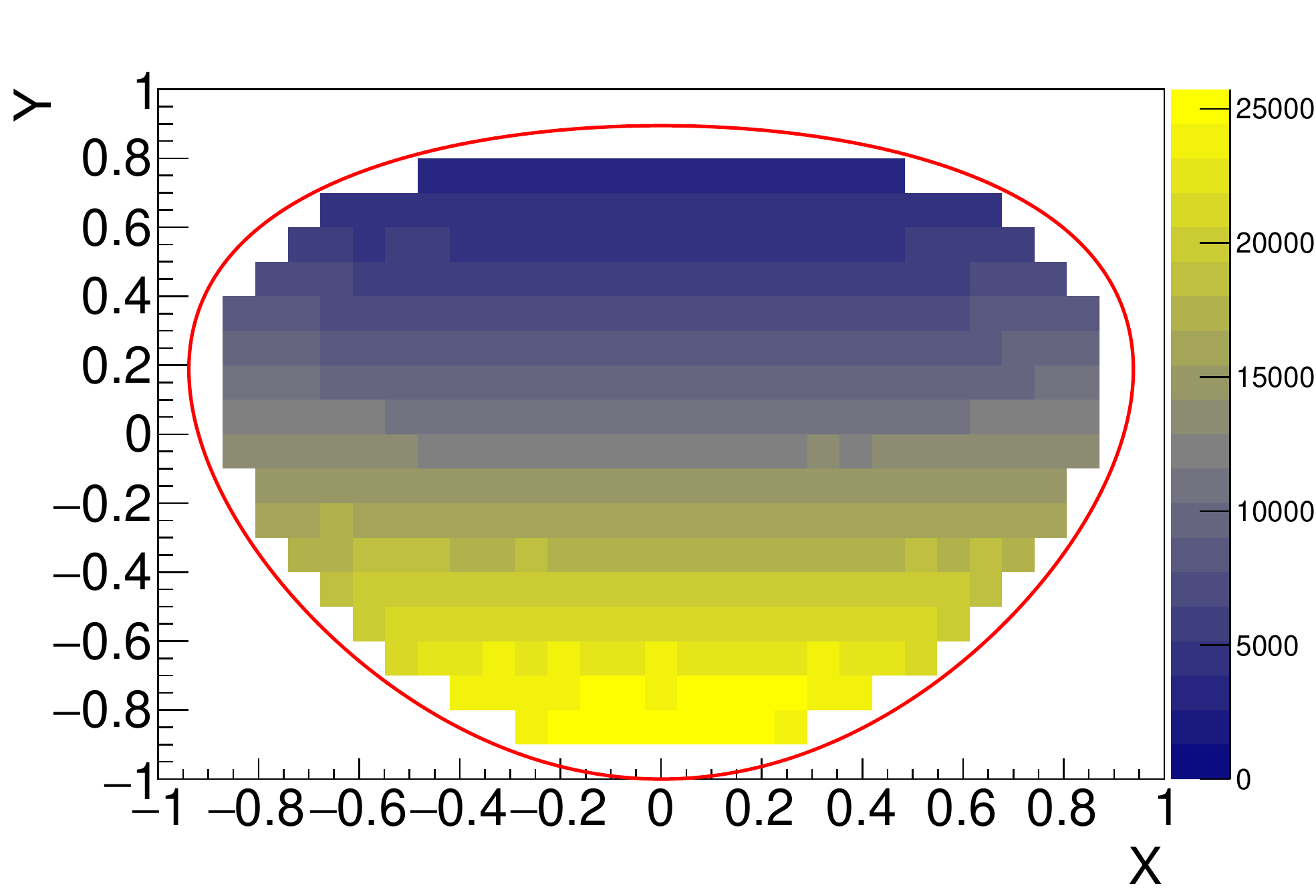}
  \caption{Left: Dalitz plot of $\eta \rightarrow \pi^+ \pi^- \pi^0$.}
  \label{fig:dalitz}
\end{figure} 

\begin{table}[!]
  \centering
  \resizebox{\textwidth}{!}{\begin{tabular}{|c|c|c|c|c|c|}
  \hline
   & a & b & d & f & g \\
  \hline
  KLOE('16)~\cite{KLOEdalitz_16} & $−1.095 \pm 0.004$ & $0.145 \pm 0.006$ & $0.081 \pm 0.007$ & $0.141 \pm 0.011$ & $−0.044 \pm 0.016$ \\
  KLOE('16)~\cite{KLOEdalitz_16} & $−1.104 \pm 0.004$ & $0.142 \pm 0.006$ & $0.073 \pm 0.005$ & $0.154 \pm 0.008$ &\\
  KLOE('08)~\cite{KLOEdalitz_08} & $−1.090 \pm 0.020$ & $0.124 \pm 0.012$ &  $0.057 \pm 0.017$ &  $0.14 \pm 0.02$ & \\
  WASA~\cite{wasa_dalitz} & $−1.144 \pm 0.018$ & $0.219 \pm 0.051$  & $0.086 \pm 0.023$ &  $0.115 \pm 0.037$ & \\
  BESIII~\cite{bes_dalitz} & $−0.128 \pm 0.017$ & $0.153 \pm 0.017$ & $0.085 \pm 0.018$ & $0.173 \pm 0.035$ &\\
  \hline
  \end{tabular}}
  \caption{Dalitz plot parameters of KLOE('16) compared to previous measurements.}
  \label{table:dalitz}
\end{table}

The Dalitz plot distribution is expressed in terms of the normalized variables: $X = \sqrt{3} \frac{T_+−T_−} {Q_{\eta}} $, $Y = 3 \frac{T_0}{Q_{\eta}} - 1$ where the $T_i$ are the kinetic energies of the pions in the $\eta$ rest frame and $Q_{\eta} = m_{\eta}-2m_{\pi^{\pm}}-m_{\pi^0}$. The squared amplitude is usually parametrized as a Taylor expansion around the center, $|A(X;Y)|^2 \simeq N(1+ aY + bY^2 + cX + dX^2 + eXY + fY^3 + gX^2Y + ...)$. The Dalitz plot has been recently measured by KLOE-2 with the decay $e^+e^- \rightarrow \Phi \rightarrow \eta \gamma$, with $\eta \rightarrow \pi^+\pi^-\pi^0$, on a sample of 1.6 fb$^{-1}$ of data, corresponding to $4.7\times10^6$ events~\cite{KLOEdalitz_16}, improving the previous measurement and extracting the $g$ parameter for the first time. The Dalitz plot is shown in Fig.~\ref{fig:dalitz} and a list of the parameters compared to other measurements is presented in table~\ref{table:dalitz}. The c and e parameters are C-violating and are compatible with zero.

\section{Dark Photon Searches}
\begin{figure}[htpb!]
\centering
\includegraphics[width=5cm]{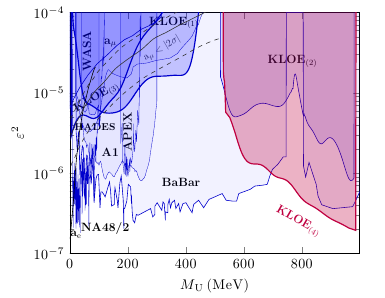}
\caption{90\% CL exclusion plot  for $\varepsilon^2$ as a function of the $\mathrm{U}$-boson mass.  
The solid lines are the limits from the muon and electron anomaly. The gray line shows the U boson parameters that could explain the discrepancy between SM prediction and the experimental value of muon anomaly, $a_{\mu}$, with a $2\, \sigma$ error band (gray dashed lines).}
\label{fig-2-1}       
\end{figure}

The KLOE-2 Collaboration searched for the signature of the dark matter interaction carrier(dark photon) by investigating three processes:\\
\textit{\bf $\phi$-Dalitz Decay:}\\
The dark photon is expected to be produced in vector to pseudoscalar meson decays producing a peak in the invariant mass distribution of the electron-positron pair over the continuum Dalitz background. 
The KLOE-2 Collaboration search for the U boson by analyzing the $\phi \rightarrow \eta \mathrm{e}^+ \mathrm{e}^-$ decay, where the $\eta$ meson is tagged by $\pi^+ \pi^- \pi^0$~\cite{KLOE_UL1} (1.5~fb$^{-1}$) and $3 \pi^0$ decays~\cite{KLOE_UL2} (1.7~fb$^{-1}$). 
The final combined limit is shown in Fig.~\ref{fig-2-1} and dubbed as KLOE$_{(1)}$.
 This limit~\cite{KLOE_UL2} rules out a wide range of U-boson parameters that could explain the $a_{\mu}$ discrepancy in the hypothesis of visibly-decaying dark photon.\\ 
\textit{\bf Higgsstrahlung process:}\\
The KLOE Collaboration investigated also the Higgsstrahlung process, sensitive to the dark coupling constant $\alpha_{\rm D}$. The invisible scenario, where the dark Higgs is lighter than the U boson and escapes detection was considered~\cite{enrico}. In this case, the expected signal is a muon pair from the U-boson decay plus missing energy.
The analysis was performed using two data samples: one collected on the $\phi$ peak and a second one at $\sqrt{s}=1000$~MeV (off-peak sample), which is not affected by resonant backgrounds.
 No signal signature has been observed and a Bayesian limit on the number of signal events at 90\% CL has been evaluated, bin-by-bin, for the on-peak and off-peak sample separately. 
In the hypothesis of $\alpha_\mathrm{D}=\alpha$, the upper  limits on the kinetic mixing parameter $\varepsilon$ ranges between $10^{-4}$ and $10^{-3}$.\\
\textit{\bf U boson radiative production:}\\
KLOE-2 investigated both leptonic and hadronic decays of the U boson radiately produced in $e^+e^-$ annihilation and decaying into: $ \rm U \to \mu^+ \mu^-,\, \pi^+\pi^-$ and  $\rm U\to \mathrm{e^+ e^-}$. 

No significant dark photon signature has been observed and limits at 90\% CL have been extracted for all processes on the number of U events.
The limits on U events have been converted in limits on the kinetic mixing parameter $\varepsilon^2$ by using the formula reported in Refs.~\cite{eeg,mmg,ppg}. These limits are shown in Fig.~1 (KLOE$_{(2,3,4)}$).

\section{Discrete symmetries and quantum decoherence tests}

Pairs of entangled neutral kaons, produced at DA$\Phi$NE from the $\Phi$ meson decay, represent an unique system to perform precise tests of discrete symmetries and Quantum Mechanics principles. 

KLOE-2 obtained the most precise results on the CPT-violating parameter $\delta$,  as well as Standard Model Extension(SME) coupling parameters~\cite{Kostelecky:2001ff}, based on the analysis of $\phi\to\Ks\Kl\to(\pi^+\pi^-)(\pi^+\pi^-)$ process.
In the SME framework  the CPT violation manifests itself indirectly via Lorentz symmetry violation effects, namely the observables become dependent on the meson momentum with respect to the distant stars. 
The KLOE-2 results reach the sensitivity at the level of $10^{-18}$ GeV~\cite{Babusci:2013gda}, which is several orders of magnitude more precise than results obtained with other neutral meson systems~\footnote{This 
precision comes close to the Planck scale of $10^{-19}$ GeV.}. The results were
obtained with the data sample corresponding to the 1.7 fb$^{-1}$ integrated luminosity, and is mainly limited by the statistical errors.

The relations between the charge asymmetries in the semileptonic decays of $\Kl$ and $\Ks$ can be expressed in terms of 
the CP- and CPT-violating parameters. 
The details of the current analysis status can be found in Ref.~\cite{daria}.

The CPT symmetry as well as T symmetry can be also tested at KLOE by a new method in kaon transitions  
based on a comparison between rates of processes from given kaon flavor-defined to given  CP-defined states and their time-reversal conjugates obtained by an exchange of initial and final states~\cite{Bernabeu-t,Bernabeu-cpt}.  The details of the current analysis status can be found in Ref.~\cite{alek}.

Finally, the tests of quantum decoherence by    
interferometric studies of the double decay rate for a pair of neutral kaons decaying into $\ppm\ppm$ channel was performed at KLOE~\cite{decoherence}. 
The current result is limited by the statistics. 


\section{Summary and Outlook}

The KLOE-2 experiment reforms the rich scientific program including interferometry and discrete symmetry tests with kaons, $\gamma\gamma$ physics, light meson spectroscopy, dark matter searches and
hadronic physics below 1 GeV. Most of the already  presented results are limited by the statistical errors. The ongoing data taking campaign is aiming to gather the data sample of at least 5 fb$^{-1}$. The 
new analyses will take advantage not only of the larger statistics but of to the new detectors, which are expected to improve significantly the key experimental parameters e.g. time resolution, or acceptance coverage.

\section*{Acknowledgments}
We warmly thank our former KLOE colleagues for the access to the data collected during the KLOE data taking campaign.
We thank the DA$\Phi$NE team for their efforts in maintaining low background running conditions and their collaboration during all data taking. We want to thank our technical staff: 
G.F. Fortugno and F. Sborzacchi for their dedication in ensuring efficient operation of the KLOE computing facilities; 
M. Anelli for his continuous attention to the gas system and detector safety; 
A. Balla, M. Gatta, G. Corradi and G. Papalino for electronics maintenance; 
C. Piscitelli for his help during major maintenance periods. 
This work was supported in part 
by the Polish National Science Centre through the Grants No.\
2013/08/M/ST2/00323,
2013/11/B/ST2/04245,
2014/14/E/ST2/00262,
2014/12/S/ST2/00459,
2016/21/N/ST2/01727,
2016/23/N/ST2/01293.

\end{document}